\documentclass[prl,twocolumn,showpacs,superscriptaddress,amssymb,10pt]{revtex4-1}

\usepackage{graphicx}
\usepackage{dcolumn}
\usepackage{bm}
\usepackage{epsfig}
\usepackage{color}
\usepackage{longtable}
\usepackage{amsmath}
\usepackage{multirow}
\usepackage{tabularx}
\usepackage{siunitx}
\definecolor{aogreen}{rgb}{0.0, 0.5, 0.0}

\def\ketm#1{  \left\vert  #1   \right\rangle   }

\def\memred#1#2#3{  \left\langle #1 \vert\vert  #2 \vert\vert #3 \right\rangle   }

\definecolor{mymainmessagecolor}{RGB}{10,200,10}
\definecolor{revisedcolor}{RGB}{0,100,20}
\DeclareMathOperator*{\SumInt}{%
\mathchoice%
  {\ooalign{$\displaystyle\sum$\cr\hidewidth$\displaystyle\int$\hidewidth\cr}}
  {\ooalign{\raisebox{.14\height}{\scalebox{.7}{$\textstyle\sum$}}\cr\hidewidth$\textstyle\int$\hidewidth\cr}}
  {\ooalign{\raisebox{.2\height}{\scalebox{.6}{$\scriptstyle\sum$}}\cr$\scriptstyle\int$\cr}}
  {\ooalign{\raisebox{.2\height}{\scalebox{.6}{$\scriptstyle\sum$}}\cr$\scriptstyle\int$\cr}}
}

\begin{document}
\include{Bibliography.bib}
\preprint{}
\title{
Fluorescence polarization as a precise tool for understanding nonsequential many-photon ionization
}

\author{J.~Hofbrucker}
\affiliation{Helmholtz-Institut Jena, Fr\"o{}belstieg 3, D-07743 Jena, Germany}
\affiliation{Theoretisch-Physikalisches Institut, Friedrich-Schiller-Universit\"at Jena, Max-Wien-Platz 1, D-07743
Jena, Germany}%
\affiliation{GSI Helmholtzzentrum f\"ur Schwerionenforschung GmbH, Planckstrasse 1, D-64291 Darmstadt, Germany}

\author{A.~V.~Volotka}
\affiliation{Helmholtz-Institut Jena, Fr\"o{}belstieg 3, D-07743 Jena, Germany}%
\affiliation{GSI Helmholtzzentrum f\"ur Schwerionenforschung GmbH, Planckstrasse 1, D-64291 Darmstadt, Germany}

\author{S.~Fritzsche}
\affiliation{Helmholtz-Institut Jena, Fr\"o{}belstieg 3, D-07743 Jena, Germany}%
\affiliation{Theoretisch-Physikalisches Institut, Friedrich-Schiller-Universit\"at Jena, Max-Wien-Platz 1, D-07743
Jena, Germany}
\affiliation{GSI Helmholtzzentrum f\"ur Schwerionenforschung GmbH, Planckstrasse 1, D-64291 Darmstadt, Germany}

\date{\today \\[0.3cm]}

\begin{abstract}
Nonsequential two-photon ionization of inner-shell $np$ subshell of neutral atoms by circularly polarized light is investigated. Detection of subsequent fluorescence as a signature of the process is proposed and the dependence of fluorescence degree of polarization on incident photon beam energy is studied. It is generally expected that the degree of polarization remains approximately constant, except when the beam energy is tuned to an intermediate $n's$ resonance. However, strong unexpected change in the polarization degree is discovered for nonsequential two-photon ionization at specific incident beam energy due to a zero contribution of the otherwise dominant ionization channel. Polarization degree of the fluorescence depends less on the beam parameters and its measurements at this specific beam energy, whose position is very sensitive to the details of the employed theory, are highly desirable for evaluation of theoretical calculations of nonlinear ionization at hitherto unreachable accuracy.
\end{abstract}

\newpage
\maketitle


\textit{Introduction.}
The advances and increasing number of free-electron lasers (FELs) revolutionized exploration of inner-shell electron dynamics and nonlinear light-matter interaction by delivering intense extreme-ultraviolet and x-ray beams. These high intensities allow detection of nonsequential ionization where multiple photons are absorbed at the same time. First observation of the lowest order nonsequential ionization, the two-photon ionization, by intense FEL beams were made by measuring high charge states of rare gas atoms, however, only ion yields in arbitrary units were extracted \cite{Sorokin:2007:213002, Richter:2009:163002}. More recent experiments do report a total ionization cross section which is extracted from measurements of ion or fluorescence yields, however, more often than not, the corresponding uncertainty cannot be estimated. While some of these experimental cross sections seem to be in agreement with theory \cite{Zernik:1964:A51, Novikov:2001:4857, Sytcheva:2012:023414, Mouloudakis:2018:053413} within an order of magnitude \cite{Tamasaku:2014:10.1038, Tamasaku:2018:083901}, other \cite{Doumy:2011:083002, Royce:2018:112} report values of even two or three orders of magnitude larger than theoretically calculated predictions. There are few measurements which were reported with the experimental uncertainties \cite{Richter:2010:194005, Ghimire:2016:043418, Szlachetko:2016:33292}, however, since second-order processes depend quadratically on the incoming beam intensity, the uncertainties in measured cross sections are strongly affected by beam parameters which determine the absolute intensity. As a consequence the uncertainty in resulting cross section can reach up to an order of magnitude. These large uncertainties smear out the differences between different theoretical models \cite{Saenz:1999:5629, Florescu:2011:033425, Bray:2012:135, Douguet:2016:033402, Grum-Grzhimailo:2016:063418, Takeshi:2016:023405, Karamatskou:2017:013415, Hofbrucker:2017:013409} preventing detailed understanding of the process as well as precise comparison between experiment and theory.

In this paper, we demonstrate that two-photon inner-shell ionization can be precisely understood by studying the degree of polarization of the subsequent fluorescent light and its dependence on incident beam energy. Firstly, degree of polarization does not depend on some beam parameters such as peak intensity, and hence the corresponding uncertainties do not propagate to the final observable quantity. Secondly, we demonstrate that in contrast to the expected near constant value of degree of polarization of the fluorescent light, clear peaks (or troughs) appear for specific nonsequential incident photon energies. The origin of the peaks is similar to the Cooper minimum in one-photon ionization process \cite{Cooper:1962:681, Cooper:1964:762}, which describes the incident photon energy at which the dominant ionization channel vanishes. This behavior is strongly projected into observable quantities such as total cross section and photoelectron angular distributions. The aim of this paper is to clearly prove the existence of such Cooper minima in nonsequential two-photon ionization and present their strong influence on the degree of polarization of subsequent fluorescent light. Measurement of fluorescence as a characteristic of two-photon ionization is well established \cite{Tamasaku:2014:10.1038, Ghimire:2016:043418, Szlachetko:2016:33292}, and hence, additional detection of fluorescence polarization degree could be utilized to evaluate many-electron or strong field effects. Moreover, this method could find applications beyond fundamental importance, as it could also serve for applied fields such as nonlinear spectroscopy of atoms and molecules \cite{Tamasaku:2018:083901}.

For clarity, but without loss of generality, let us explain the suggested principle on an example which can be schematically represented as follows
\begin{equation}
   \textrm{A}+2\gamma_i \rightarrow \textrm{A}^{+}np^{-1}+ e \rightarrow \textrm{A}^{+}n's^{-1} + e + \gamma_f .
\end{equation}
In this two-step process, inner-shell $np^{-1}$ vacancy is created in an atom A by absorption of two right-circularly polarized photons, each with energy $\omega_i$. This vacancy is subsequently filled by a radiative decay from energetically higher $n's$ state resulting in an emission of a fluorescent photon. Our goal is to analyze degree of polarization of this photon. First, we start with simple and clear explanation of the polarization properties. For this reason we will restrict our description to nonrelativistic single-active electron framework, where the interaction of the active electron with the field is considered within electric dipole approximation. Later, however, we justify the obtained predictions within fully relativistic many-body description and confirm them numerically on an example of magnesium atom.

\textit{Nonrelativistic explanation of nonlinear Cooper minimum.}
To describe the interaction of the atom with light, we have previously developed a theoretical approach \cite{Hofbrucker:2016:063412}. Tracing out degrees of freedom of the photoelectron leaves us with density matrix $\rho_{\text{ion}}^{m m^{'}}$ which describes the final photo-ion with an inner-shell hole in the magnetic substate $m$ of the $p$-orbital
\begin{eqnarray}\label{Eq.DensityMatrixIon}
 \rho_{\textrm{ion}}^{11}&=&\frac{9}{140\pi^2}27|U_{f}|^2\nonumber,\\
 \rho_{\textrm{ion}}^{00}&=&\frac{9}{140\pi^2}9|U_{f}|^2,\\\nonumber
 \rho_{\textrm{ion}}^{-1-1}&=&\frac{9}{1400\pi^2}(7|U_{p}|^2+18|U_{f}|^2),\\\nonumber
 \rho_{\textrm{ion}}^{m m^{'}}&=&0, \hspace{1cm}\textrm{for all $m \neq m^{'}$}.
\end{eqnarray}
Here, $U_l$ are the radial transition amplitudes to the final $l-$partial wave which contain the summation over the complete spectrum of intermediate ionization paths 
\begin{equation}\label{Eq.TransitionAmplitude}
    U_l=\SumInt_i\frac{\memred{\epsilon_e  l}{\textbf{r}}{i}\memred{i}{\textbf{r}}{np}}{E_{np}+\omega-E_i},
\end{equation}
where typical notation for quantum numbers was used. The contributions to the final $p$ partial wave originate from two quantum paths, $p\rightarrow s \rightarrow p$ and $p\rightarrow d \rightarrow p$, and its corresponding amplitude is given by $U_{p}=(5U_{p\rightarrow s\rightarrow p}+U_{p\rightarrow d\rightarrow p})$.

To obtain polarization properties of the $n's\rightarrow np$ decay, we once again use tools of density matrix formalism. The general form of fluorescent photon density matrix was derived, e.g., in Ref. \cite{Sharma:2010:023419}, and in our case it is given by
\begin{figure}
    \centering
    \includegraphics[scale=0.25]{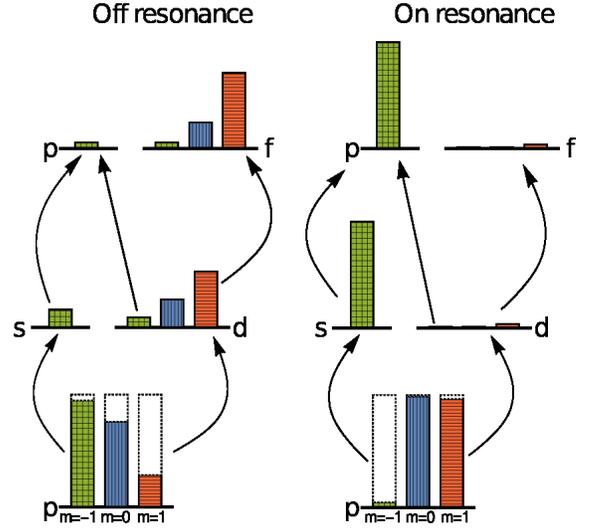}
    \caption{(Color online) Schematic diagram of possible electric dipole channels for two-photon ionization of a $p-$electron by right-circularly polarized light. The population of each magnetic projection, $m=-1$ (squared green pattern), $m=0$ (vertical blue pattern), $m=1$ (horizontal red pattern), after ionization is indicated with respect to the full shell population (dotted rectangle) for photon energies not matching polarization resonances (left) and matching them (right).}
    \label{fig:Channels}
\end{figure}
\begin{eqnarray}\label{Eq.DensityMatrixPhoton}
\rho_{\gamma_f}^{\pm1\pm1}&=& \frac{9|T_{n'snp}|^2}{2800\pi}\Big\{\nonumber
7(\cos \theta\pm1)^2 |U_{p}|^2 \\ 
&+& 12\big[(3\cos\theta\mp 7)^2-10\big] |U_{f}|^2\Big\}, \\ \nonumber
\rho_{\gamma_f}^{-11}&=&\rho_{\gamma_f}^{1-1}=\frac{-9 |T_{n'snp}|^2 \sin^2 \theta}{2800\pi}(7|U_{p}|^2+108|U_{f}|^2),
\end{eqnarray}
where $\theta$ is angle between fluorescence emission direction and propagation direction of the two incident photons, $T_{n'snp}$ is the reduced transition amplitude describing decay of the $n's$ electron to the $np^{-1}$ vacancy. The degrees of linear and circular polarization can be easily obtained from the photon density matrix
\begin{eqnarray}\label{Eq.Stokes} ~ \\\nonumber
P_1 (\theta)&=&\frac{-2 \sin^2 (\theta)\Big[7|U_{p}|^2+108|U_{f}|^2\Big]}{7\big[3+\cos (2\theta)\big]|U_{p}|^2+36\big[29+3\cos(2\theta)\big]|U_{f}|^2},\\\nonumber
P_3 (\theta)&=&\frac{28 \cos (\theta)\Big[|U_{p}|^2-36|U_{f}|^2\Big]}{7\big[3+\cos (2\theta)\big]|U_{p}|^2+36\big[29+3\cos(2\theta)\big]|U_{f}|^2}.
\end{eqnarray}
From Eq. (\ref{Eq.DensityMatrixPhoton}), it is clear that $P_2(\theta)=0$. The above expressions show, that measuring polarization near incident photon propagation direction ($\theta=\ang{0}$) will yield the largest effect for $P_3$, while placing the detector perpendicular to this axis will result in maximum of $P_1$. Moreover, it also indicates that whenever the ionization path with intermediate $s$-state strongly dominates, i.e. $|U_p|\gg|U_f|$, fluorescence polarization will be fully polarized, $P_1^2(\theta) + P_3^2(\theta)\approx 1$, with $P_3(\ang{0})\approx 1$ and $P_1(\ang{90})\approx -1$ (right side of Fig. \ref{fig:Channels}). However, according to the Fano propensity rule \cite{Fano:1969:131}, the channel with the highest angular momentum generally dominates, $|U_f|\gg|U_p|$. In this case, Eqs.~(\ref{Eq.Stokes}) predict the fluorescence to have $P_3(\ang{0}) \approx -0.88$ and $P_1(\ang{90}) \approx -0.23$ (left side of Fig. \ref{fig:Channels}). We can therefore expect that any variation of the amplitude ratios will be sensitively projected into the fluorescence polarization. The same could be expected in angular distributions of fluorescent photon, but here we restrict ourselves to its polarization properties only.


Although the transition amplitudes $U_l$ need to be calculated explicitly, let us predict full polarization transfer from incident to fluorescent photon in direct two-photon ionization by circularly polarized light by physical arguments based on properties of Eq.~(\ref{Eq.TransitionAmplitude}) and quantum paths of each magnetic substate of the $p$ orbital (see Fig. \ref{fig:Channels}). The absorption of each right-circularly polarized photon increases angular momentum projection of the active electron by unity, which restricts the possible ionization channels for each initial magnetic substate. While the ionization channel which describes the photoelectron with $f-$symmetry is open to all of the bound $np-$electrons, channels describing photoelectron with $p-$symmetry is open only to $m=-1$ electron. It follows that whenever $U_p$ transition amplitude dominates over the $U_f$ amplitude, ionization of the electron with negative projection will be strongly preferred. If we, for example, choose a photon energy matching a transition to an intermediate level $s-$resonance, channel with final $p-$symmetry will be strongly enhanced and population of negative projection electrons will be depleted (see right side of Fig. \ref{fig:Channels}). The subsequent $n's \rightarrow np$ electronic transition will therefore mostly fill the $m=-1$ vacancy, and emit a photon with helicity $\lambda=-1$ along the incident beam propagation (quantization) axis. This means that the emitted radiation will be fully polarized, and specifically, degree of circular polarization of this photon along quantization axis will be $P_3(\ang{0})\approx 1$ while at right angle to the quantization axis $P_1(\ang{90})\approx -1$. The resonant photon energy will be therefore imprinted in fluorescence polarization spectrum in form of a peak (or through), to which we shall in future refer to as \textit{level resonance}. Since the channel with final $f-$symmetry is generally dominant, tuning our photon energy to an intermediate $d-$resonance will not affect the fluorescence polarization.

There is another scenario, where transition amplitude $U_p$ dominates over $U_f$ amplitude. Equation~(\ref{Eq.TransitionAmplitude}) shows, that all possible intermediate ionization paths contribute to the transition amplitude, and hence, we need to sum over the complete atomic spectrum. Now, consider a photon energy $\omega$ for which $E_{np}+\omega-E_{d_1}<0$ and $E_{np}+\omega-E_{d_2}>0$. We can therefore always \textit{fine tune} the incident photon energy, such that contributions to the amplitudes from higher energy paths $d_1$ and from lower energy spectra $d_2$
\begin{eqnarray}\label{Eq.Bips} \nonumber
   U_f(\omega)&=&\SumInt_{d_1}^{E_{d_1}>E_{np}+\omega}\frac{\memred{\epsilon_e  f}{\textbf{r}}{d_1}\memred{d_1}{\textbf{r}}{np}}{E_{np}+\omega-E_{d_1}}\\
   &+&
    \SumInt_{d_2}^{E_{d_2}<E_{np}+\omega}\frac{\memred{\epsilon_e  f}{\textbf{r}}{d_2}\memred{d_2}{\textbf{r}}{np}}{E_{np}+\omega-E_{d_2}},
\end{eqnarray}
exactly balance each other out, i.e. $U_f(\omega_{\textrm{NCM}})=0$. This energy describes passing of the otherwise dominant channel through zero, and hence, we call this point the nonlinear Cooper minimum (NCM). At this incident energy, only electrons with initial $m=-1$ projection will be ionized and hence the subsequent fluorescence will be fully polarized. Such nonlinear Cooper minimum can be found between any two adjacent level resonances of the same angular momentum. Furthermore, this is a general feature of two-(or even multi-)photon ionization. If an initially bound electron in $nl$ shell absorbs multiple photons, nonlinear Cooper minimum can be found between any pair of $n'(l+1)$ and $(n'+1)(l+1)$ resonances. In our example, switching off the channels with intermediate $d-$ states results in a polarization resonance, which we coin \textit{Cooper resonance}.

\textit{Nonsequential two-photon photon ionization of magnesium atom.}
Let us support the above predictions with a numerical calculation of two-photon ionization of $p$ electron of neutral magnesium. We will use the typical many-electron notation in which $J$ represents the total angular momentum, $M$ its projection and $\alpha$ all further numbers that are needed for unique characterization of the state. An atom being initially in the state $\ketm{\alpha_i J_i M_i}$ is ionized by two right-circularly polarized photons $\ketm{\bm{k}_i \lambda_i}$, each with energy $\omega_i$. After simultaneous absorption of two photons, the system composes of a singly charged excited ion and a free electron with momentum $\textbf{p}_e$ and spin projection $m_e$, i.e. final state after ionization $\ketm{\alpha_f J_f M_f, \textbf{p}_e m_e}$. In the next step, the excited ion relaxes into its ground state with simultaneous emission of a fluorescent photon $\ketm{\textbf{k}_{f}\lambda_f}$ with momentum $\textbf{k}_{f}$, helicity $\lambda_f$ and energy $\omega_f$.

\begin{figure*}[t]
    \centering
    \includegraphics[scale=0.65]{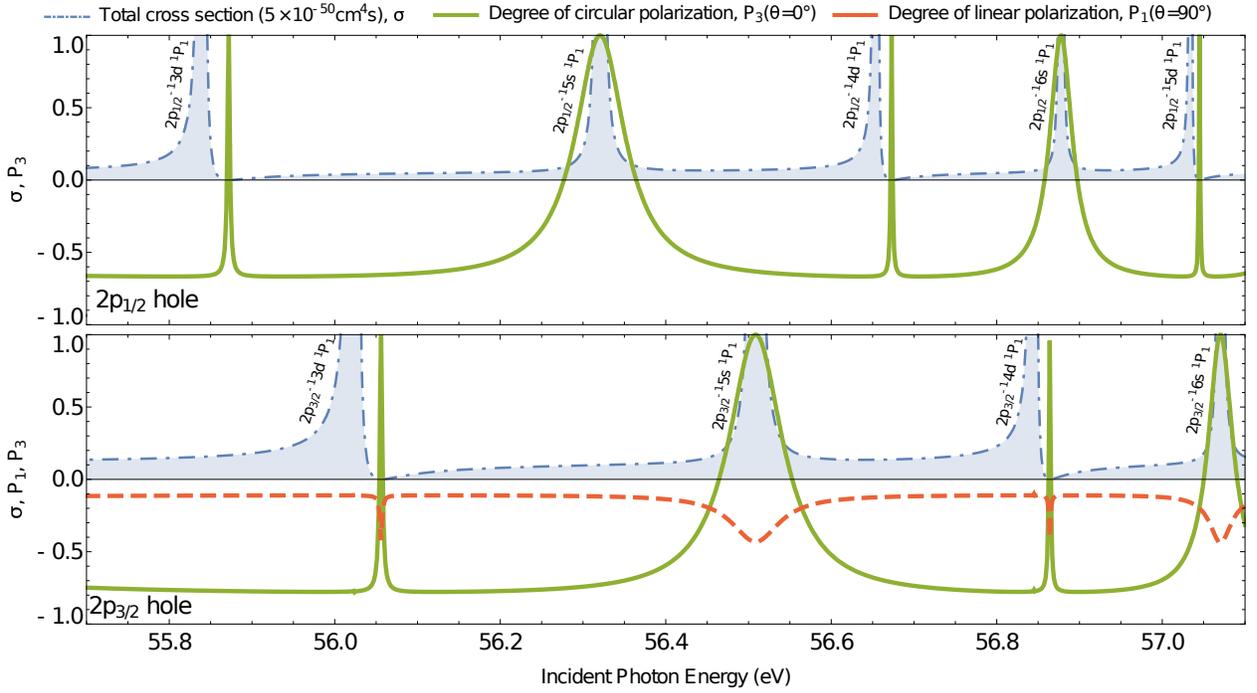}
    \caption{(Color online) Two-photon ionization of the $p_{1/2}-$(upper) and $p_{3/2}-$ (lower) electron of magnesium atom. Total cross section (dot-dashed) and degree of circular and linear 
    polarization of the emitted fluorescent photon $P_3(\theta=\ang{0})$ (full) and $P_1(\theta=\ang{90})$ (dashed) as functions of incident photon energy. While the level fluorescence resonances match the intermediate $s-$state (also visible in the total cross section), the Cooper resonances originate from vanishing of the dominant ionization channels.}
    \label{Fig:CrossSectionsP3}
\end{figure*}
For the particular example of two-photon ionization of the $p$ orbital of magnesium atom, the initial state is $\ketm{\alpha_i J_i M_i} = \ketm{1s^2 2s^2 2p^6 3s^2 ~ ^1S_0} \equiv i$. The dominant ionization channels will lead to the following two final states (ionic part only) $\ketm{1s^2 2s^2 2p^5 3s^2 ~ ^2P_{1/2}} \equiv f_{1/2}$ and $\ketm{1s^2 2s^2 2p^5 3s^2 ~ ^2P_{3/2}} \equiv f_{3/2}$, describing singly charged magnesium ion with $2p_{1/2}^{-1}$ and $2p_{3/2}^{-1}$ holes, respectively. The subsequent decay of the ion to the ground state 
$\ketm{1s^2 2s^2 2p^6 3s ~ ^2S_{1/2}} \equiv g$ results in an emission of a fluorescence photon with energy $\omega_f = E_{f_{1/2}} - E_g = 50.153$ eV or $\omega_f = E_{f_{3/2}} - E_g = 49.915$ eV \cite{NIST}. Let us now consider displacement of other electrons, so called shake-up processes, which could influence the final result. Assuming that this secondary effect is more important near threshold energies, similarly as in one-photon ionization, the shake-up transition $1s^2 2s^2 2p^6 3s^2 + 2\gamma_i \rightarrow 1s^2 2s^2 2p^5 3s 4s$ dominates this effect \cite{Badnell/JPB:1997}. We estimated that in our example, this process  contributes by less than 10\%. Moreover, the subsequent fluorescent photon carries energy  $E_{1s^2 2s^2 2p^5 3s 4s} - E_{1s^2 2s^2 2p^6 4s} \geq 52.26$ eV \cite{NIST}, depending on coupling of $2p^53s4s$ electrons. It is therefore clear, that in comparison to $\approx 50$ eV produced by the dominant channels, one can filter out the photons arising from the shake-up decay channel by the photon detector \cite{Wunsche/RSI:2019}.

Based on the above arguments, it is reasonable to consider the two-photon ionization process which leads to the final states $f_{1/2}$ and $f_{3/2}$ only. The energy of the subsequent fluorescent photons emitted by a decay of these two states differs by about $0.25$ eV, which is  well within experimentally resolvable limits \cite{Wunsche/RSI:2019}. Therefore, these photons can be easily distinguished and their density matrix is given by
\begin{eqnarray}\label{Eq.PhotonDensityMatrix}
     \rho^{\lambda_f \lambda_f'}_{\gamma_f} = \sum_{\substack{M_g M_g'\\M_f M_f'}}
     T_{f_j M_f g M_g}^{\lambda_{f}} \rho_{\textrm{ion}}^{f_j M_f f_j M_f'}
     T_{f_j M_f' g M_g'}^{\lambda_{f}'*} ,
\end{eqnarray}
where $T_{f_j M_f g M_g}^{\lambda_{f}}$ is the transition amplitude describing decay of the singly charged ion in the state $f_j$ to its ground state $g$, and $\rho_{\textrm{ion}}^{f_j M_f f_j M_f'}$ is the density matrix of the ion \cite{Hofbrucker/PRA:2016}
\begin{eqnarray}\label{Eq.IonDensityMatrix}
\rho_{\textrm{ion}}^{f_j M_f f_j M_f'} = \frac{(8 \pi)^3 \alpha^2}{\omega_i^2} \int d^3\bm{p}_e
\sum_{\substack{\lambda_{i_1} \lambda_{i_2} m_e\\ \lambda_{i_1}' \lambda_{i_2}' m_e'}} 
\rho_{\gamma_i}^{\lambda_{i_1} \lambda_{i_1}'}
\rho_{\gamma_i}^{\lambda_{i_2} \lambda_{i_2}'}\nonumber\\
\times
M_{i f_j M_f \bm{p}_e m_e}^{\lambda_{i_1} \lambda_{i_2}}
M_{i f_j M_f' \bm{p}_e m_e'}^{\lambda_{i_1}' \lambda_{i_2}'*}
\delta(E_i+2\omega_i-E_{f_j}-\varepsilon_e) .
\end{eqnarray}
The transition amplitudes take the form
\begin{eqnarray}
\label{Eq.TransitionAmplitudeMB}
M_{i f_j M_f \bm{p}_e m_e}^{\lambda_1 \lambda_2} =~\int\kern-2.5em\sum_{\alpha_n J_n M_n}
   \frac{T_{\alpha_{n}J_{n}M_{n} f_j M_f \textbf{p}_e m_e}^{\lambda_{i_2}}
	     T_{i \alpha_{n}J_{n}M_{n}}^{\lambda_{i_1}}}
        {E_{i}+\omega_i-E_{\alpha_n J_n}} .
\end{eqnarray}
Unfortunately, so far there is currently no approach developed to calculate the second-order matrix elements of Eq.~(\ref{Eq.TransitionAmplitudeMB}) within many-body approaches. For this reason, we perform fully relativistic calculation within single-active electron approach. The results are presented in Fig.~\ref{Fig:CrossSectionsP3}, where the upper figure shows the results corresponding to two-photon ionization of the $2p_{1/2}$ ($f_{1/2}$ state) electron and the figure below to ionization of the $2p_{3/2}$ ($f_{3/2}$ state) electron. Figure~\ref{Fig:CrossSectionsP3} shows the total two-photon ionization cross section as well as degrees of circular and linear polarization of subsequent fluorescent photon as functions of incident photon energy. The degree of circular polarization (full green) generally takes constant value of around $P_3 (\ang{0})\approx-0.7$ along the quantization axis, while degree of linear polarization at perpendicular direction is $P_1(\ang{90})\approx-0.1$ for the $2p_{3/2}$ hole case. However, these values change dramatically at clearly visible resonances. Comparing the energy positions of these polarization resonances, we notice that some of them (level resonances) appear at the same energies as the resonances of the total cross section. These polarization resonances correspond to the sequential (resonant) two-photon ionization case. On the other hand, Cooper (polarization) resonances have no counter part in the total cross section. They appear due to the dominant ionization channel passing through zero between two adjacent $2p^6 3s^2~^1S_0 \rightarrow 2p^5 (P_{1/2}) 3s^2 nd~^1P_{1}$ or $2p^6 3s^2~^1S_0 \rightarrow 2p^5 (P_{3/2}) 3s^2 nd~^1P_{1}$ resonances, as has been explained above.

Comparing the presented results with the nonrelativistic predictions, we notice that by resolving the fine-structure splitting of the $2p-$electrons, we also observe the splitting of the Cooper resonances associated with each of them. Furthermore, Fig.~\ref{Fig:CrossSectionsP3} also shows that the relativistic description keeps the nonzero linear polarization of the fluorescence photons only for $2p_{3/2}$ hole and changes its magnitude. When, however, we assume $E_{f_{1/2}} = E_{f_{3/2}}$ an additional interference term $\rho_{\textrm{ion}}^{f_{1/2} M_f f_{3/2} M_f'}$ should be included in Eq.~(\ref{Eq.PhotonDensityMatrix}). With this interference term the nonrelativistic results are restored. Otherwise, the simple nonrelativistic description presented above predicts and clearly explains the physics of the nonlinear Cooper resonances in the fluorescence polarization {\textit spectrum}.

First Cooper resonance occurs at energy $\omega = 55.87$~eV near the intermediate $2p^6 3s^2 ~^1S_0 \rightarrow 2p^5 (P_{1/2}) 3s^2 3d ~^1P_1$ resonant transition in case of the $2p_{1/2}$ hole, while the corresponding Cooper resonance in the $2p_{3/2}$ hole fluorescence can be found at $\omega = 56.06$~eV close to the intermediate $2p^6 3s^2~^1S_0 \rightarrow 2p^5 (P_{3/2}) 3s^2 4d ~^1P_1$ resonance. The required photon energy and polarization needed for experimental verification of the above predictions are already available at the FERMI FEL \cite{Allaria:2013:913}. Furthermore, FERMI's seeded operation delivers high stability of output central wavelength allowing to resolve the narrow Cooper resonances ($\approx 0.02$~eV in the case of magnesium, see Fig.~\ref{Fig:CrossSectionsP3}). 

%
\begin{figure*}[t]
    \centering
    \includegraphics[scale=0.5]{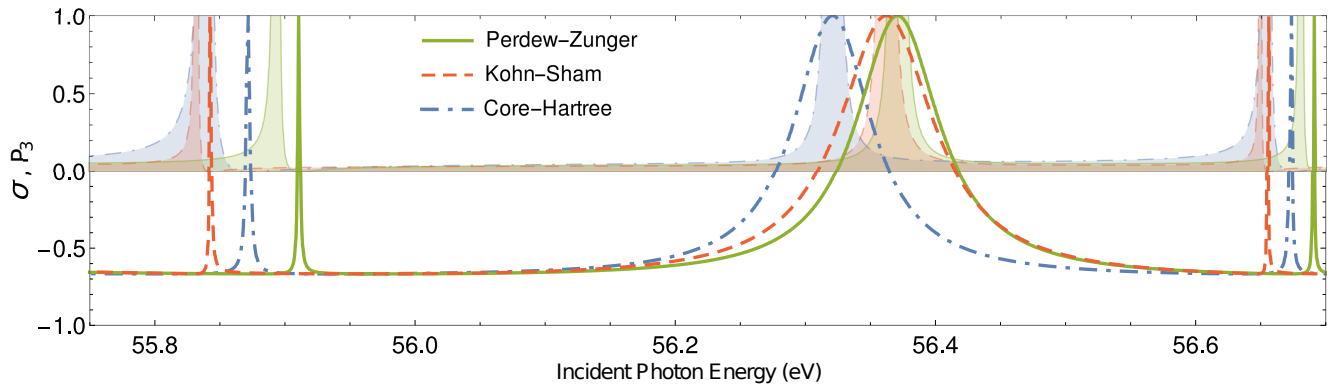}
    \caption{(Color online) Two-photon ionization of the $2p_{1/2}-$electron of magnesium atom. Total cross section (in units of $5 \times 10^{-50}$~cm$^4$s shaded) and degree of circular polarization (lines) of the emitted fluorescent photon $P_3(\theta=\ang{0})$ as functions of incident photon energy. Three different screening potentials were used to calculate the quantities: Core-Hartree (dot-dashed, blue),  Perdew-Zunger (full, green), and Kohn-Sham (dashed, red).}
    \label{Fig.Potentials}
\end{figure*}
Predictions of exact energy positions of polarization Cooper resonances strongly depend on capability of theoretical models to represent the complete atomic spectrum. However, our aim here is not to predict the position precisely, but to \textit{justify} that the nonlinear Cooper minimum always appears between any pair of $n'(l+1)$ and $(n'+1)(l+1)$ resonances (as explained above). Although we consider our proof to be rather general, we carried out further calculations for verification. The employed single-active electron approach reproduces all the intermediate resonances $\ketm{\alpha_{n}J_{n}M_{n}}$ as listed by NIST~\cite{NIST} in the given energy region. The remaining difference between single- and many-electron approaches comes from the numerical values of matrix elements. In order to investigate the sensitivity of the matrix elements on the employed theoretical model, we perform the calculations for number of different screening potentials~\cite{Perdew:1981:5048,Kohn:1965:A1133,Slater:1951:385}. Figure~\ref{Fig.Potentials} presents the cross section (shaded) and degrees of circular polarization of fluorescence light emitted after two-photon ionization of $2p_{1/2}$ electron for three different screening potentials: Core-Hartree (dot-dashed, blue), Perdew-Zunger (full, green), and Kohn-Sham (dashed, red). From the figure, it is clear that although the exact position predictions of Cooper minima differ, the minima are predicted by all potentials. Thus, with the presented calculations we cover the essential attributes of the second-order many-electron matrix elements (\ref{Eq.TransitionAmplitudeMB}) and justify the existence of nonlinear Cooper minima.

\textit{Experimental accuracy.}
The fluorescence yield $\Phi (\theta)$, which is typically measured in experiments \cite{Tamasaku:2014:10.1038, Ghimire:2016:043418, Szlachetko:2016:33292, Tamasaku:2018:083901}, being integrated over angles, is a product of cross section $\sigma^{(2)}$ and a photon flux $F$ squared \cite{Lambropoulos:1976}
\begin{equation}
   \int \Phi (\theta) d\theta \propto \sigma^{(2)} F^2.
\end{equation}
Thus, the precision of cross section extracted from a measured fluorescence yield are highly sensitive to uncertainties in the flux, which needs to be extracted from fluctuating pulse energy and pulse structure. For example, for two-photon ionization of the $K-$shell of zirconium, the total cross section was reported in the range $3.9-57\times 10^{-60}$cm$^4$s. The reason for this wide range of values was appointed to systematic uncertainty in beam parameters of LCLS FEL which determine the absolute flux \cite{Ghimire:2016:043418}. Similarly, Richter \textit{et al.} \cite{Richter:2010:194005} report $30\%$ uncertainty in the beam flux at FLASH FEL which results in $\approx 50\%$ uncertainty in the extracted cross section of two-photon ionization of singly and doubly charged neon atoms. On the other hand, the polarization degree of fluorescence is independent of these factors as it is given by a normalized difference of the measured fluorescent photon intensities. For example, the degree of linear polarization is determined by the intensities of the fluorescence light linearly polarized parallel $\Phi_{\parallel} (\theta)$ or perpendicular $\Phi_{\perp}(\theta)$ with regard to the scattering plane $P_1 (\theta) = \frac{\Phi_{\parallel}(\theta)-\Phi_{\perp}(\theta)}{\Phi_{\parallel}(\theta)+\Phi_{\perp}(\theta)}$. This insensitivity to photon flux will allow to measure the polarization with much higher accuracy than total cross section, and hence, this method could serve as an alternative approach in fundamental research as well as nonlinear spectroscopy \cite{Tamasaku:2018:083901}.

\textit{General interest.}
To illustrate the significance of nonlinear Cooper minima, we compare them to other known effects in different processes which contributed to accurate understanding of atomic processes. First of all, they can be considered as nonlinear equivalents of the Cooper minima in one-photon ionization process \cite{Cooper:1962:681, Cooper:1964:762}, since in both cases vanishing of the dominant ionization channel strongly influences observable quantities. In one-photon ionization processes, Cooper minima have opened the door for accurate comparison of theory with experiment e.g. \cite{Derenbach:1983:L337, Fahlman:1983:L485}, or even allowed to resolve relativistic effects in photoelectron angular distributions or ionization time delay \cite{Johnson:1978:1167,Hemmers:2003:053002,Saha:2014:053406,Ilchen:2018:4659}. The nonlinear Cooper minima also share similar characteristics with the tune-out wavelengths, specific photon energies at which the atomic polarizability vanishes. Their exact energy positions are sensitive even to small contributions such as finite nuclear mass effect, or relativistic and QED corrections \cite{Drake:2008:45, Henson:2015:043004, Zhang:2016:052516}. Similarly to both above examples, Cooper resonances have the potential to push the accuracy at which we understand nonsequential nonlinear process beyond current possibilities. It is worth noting that nonlinear Cooper minima are not only imprinted in the fluorescence polarization, but can also be used to find strong nonlinear interference effects such as photoelectron elliptical dichroism. It has been shown that maximum elliptical dichroism can be achieved in the case of two-photon ionization of $s-$ states for two incident photon energies near the nonlinear Cooper minima \cite{Hofbrucker:2018:053401,Manakov:2009:557}.

\textit{Conclusion.}
The existence of Cooper minima in two-photon ionization was generally proven and supported by numerical calculation. Moreover, based on physical angular momentum arguments, we demonstrated that in the case of two-photon inner-shell ionization of $p-$electrons by circularly polarized light at these nonlinear Cooper minima, the subsequent fluorescence will be unexpectedly strongly polarized. This polarization transfer from the incident to fluorescent photon appears as a strong resonance in the polarization spectra, which can be accurately measured and subsequently used as a sensitive quantity for comparison with theory. The proposed approach was demonstrated on experimentally plausible example of two-photon ionization of magnesium.
\begin{acknowledgments}
We are grateful to Markus Ilchen for enlightening discussions on current experimental possibilities. We acknowledge the support from the Bundesministerium f\"ur Bildung und Forschung (Grant No. 05K16FJA). 
\end{acknowledgments}
%
%
%
%
%
\end{document}